\documentclass[11pt]{article}
\usepackage{moriond,epsfig}
\pdfoutput=1

\def\JPsi{\ensuremath{{J}\hspace{-.08em}/\hspace{-.14em}\psi}}

\def\be{\begin{equation}}
\def\ee{\end{equation}}
\def\bea{\begin{eqnarray}}
\def\eea{\end{eqnarray}}

\def\pt {\ensuremath{p_\mathrm{T}}}
\begin{document}
\vspace*{4cm}
\title{HEAVY FLAVOR PRODUCTION AT CMS}

\author{ K. A. ULMER, on behalf of the CMS collaboration}

\address{Department of Physics, University of Colorado, \\
390 UCB, Boulder, CO, USA}

\maketitle\abstracts{
Measurements of heavy flavor production in $pp$ collisions at $\sqrt{s}$ = 7.0 TeV
recorded at the CMS experiment are presented. Double differential cross sections
with respect to transverse momentum and rapidity are shown for
\JPsi\ and $\Upsilon(1S)$, $\Upsilon(2S)$, and  $\Upsilon(3S)$. The inclusive open
beauty rate is measured with two different techniques, including a study of the 
angular correlations between $b$ jets in events with two identified $b$ jets. Lastly,
the $B^+$, $B^0$, and $B^0_s$ production rates are measured from the reconstruction
of exclusive final states.}

\section{Introduction}
Cross sections for heavy quark production in hard scattering interactions provide 
an interesting testing ground for QCD calculations. Theoretically, large
uncertainties remain due to the dependence on the renormalization and factorization
scales. Measurements from the LHC at the
center of mass energy of $\sqrt{s}$ = 7.0 TeV provide new opportunities to test and
further our understanding of the heavy quark production mechanisms.

CMS is a general purpose experiment at the Large Hadron 
Collider\,\cite{CMS}\hspace{-.35em}.
The main detector components used in these
analyses are the silicon tracker and the muon systems. 
The silicon tracker measures charged particles in the pseudorapidity
range $|\eta| < 2.5$ within a 3.8~T field of the superconducting solenoid.
It provides an impact parameter resolution of $\sim$\,15~$\mu$m and a 
\pt\ resolution of about 1.5\% for particles with transverse momenta up to
100 GeV. Muons are measured in the pseudorapidity range $|\eta|< 2.4$
with detection planes made using three technologies: drift tubes, cathode 
strip chambers, and resistive plate chambers.

\section{Onia production}

The first heavy flavor production measurements at CMS were made by
reconstructing $\JPsi$~\cite{JPsi} and $\Upsilon$~\cite{Upsilon} mesons
in their decays to two muon final states. Candidates are formed by
fitting pairs of oppositely-signed muons to a common vertex and event
yields are obtained by fitting invariant mass distributions. The
observed yields are corrected for detector acceptance, 
reconstruction inefficiencies, and trigger inefficiencies in bins
of candidate transverse momentum \pt\ and rapidity $y$ to measure
double differential cross sections. The fraction of $\JPsi$ mesons
produced from long-lived $B$ decays is also measured by fitting the
lifetime distribution of the reconstructed $\JPsi$ mesons. The three
lowest $\Upsilon$ states are all visible due to the excellent mass resolution
of the CMS detector. The yields of the $\Upsilon(2S)$ and $\Upsilon(3S)$ 
states are measured
with respect to the $\Upsilon(1S)$ state as functions of the $\Upsilon$
\pt\ and $y$.

\section{Inclusive open beauty production}

Two independent techniques are used to measure inclusive beauty production. 
The first technique makes use of semi-muonic decays of 
$B$ hadrons~\cite{inclusive}\hspace{-.35em}.
Reconstructed charged tracks with $\pt > 300$ MeV are clustered into jets
with the anti-$k_{\mathrm{T}}$ algorithm with $R = 0.5$. Events are then selected where
the jet contains a reconstructed muon with $\pt > 6$ GeV and $|\eta| < 2.1$. 
For jets originating from $b$ quarks, the decay kinematics
demand that, on average, the muon direction will be further displaced from the
jet direction than for muons from lighter jets ($udscg$). The observed distribution
of the quantity $\pt^{rel} = |\vec{p}_{\mu}\times\vec{p}_j|/|\vec{p}_{\mu}|$ is fit
to separate signal $b$ jets from background. Templates for the signal and background
$\pt^{rel}$ shapes are obtained from simulation (for signal and $c$ quark backgrounds) or
data ($uds$ quark and gluon backgrounds) and are crosschecked in data for those obtained
from simulation.

\begin{figure}
\includegraphics[angle=90,width=0.55\textwidth]{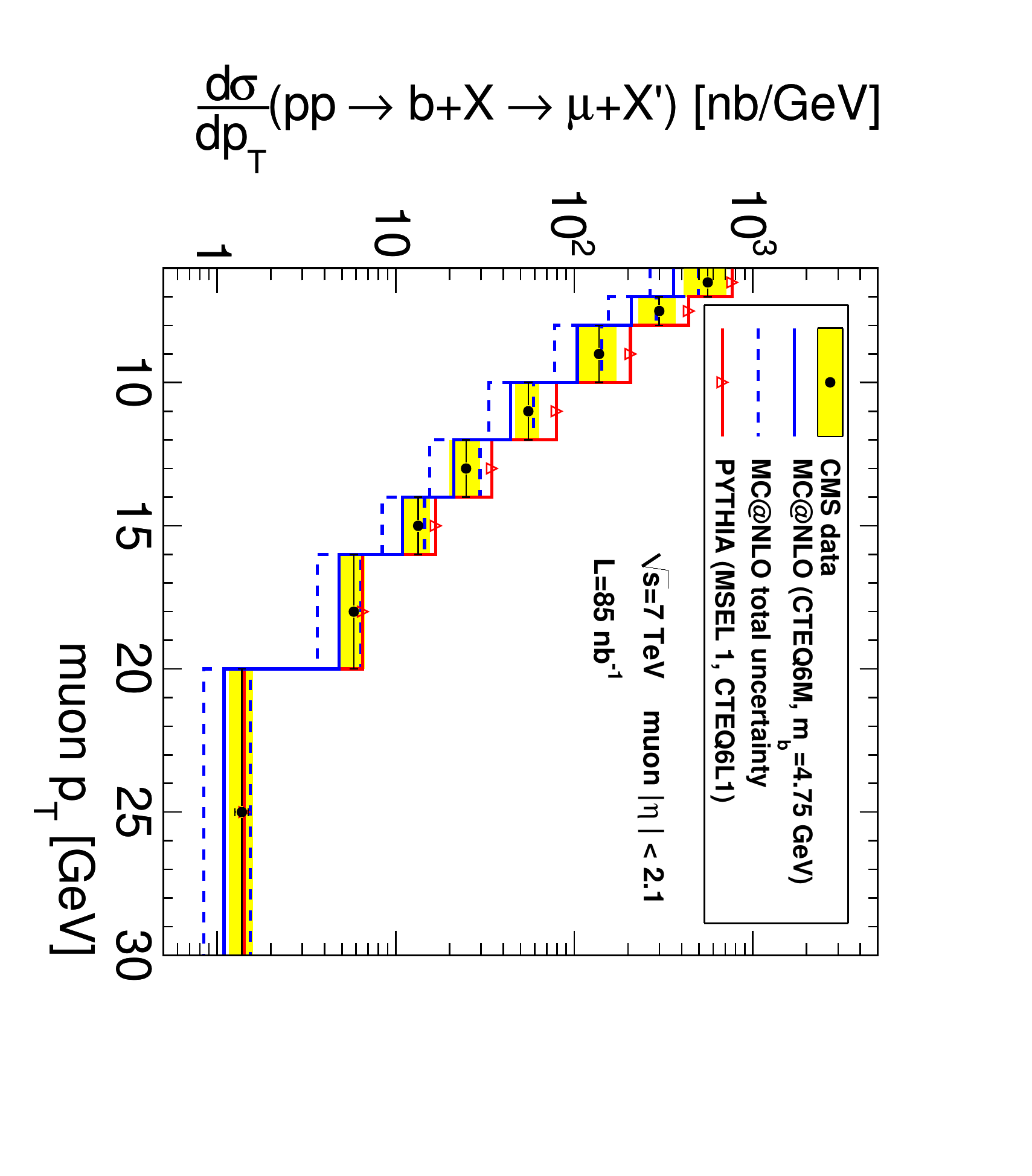}
\includegraphics[width=0.44\textwidth]{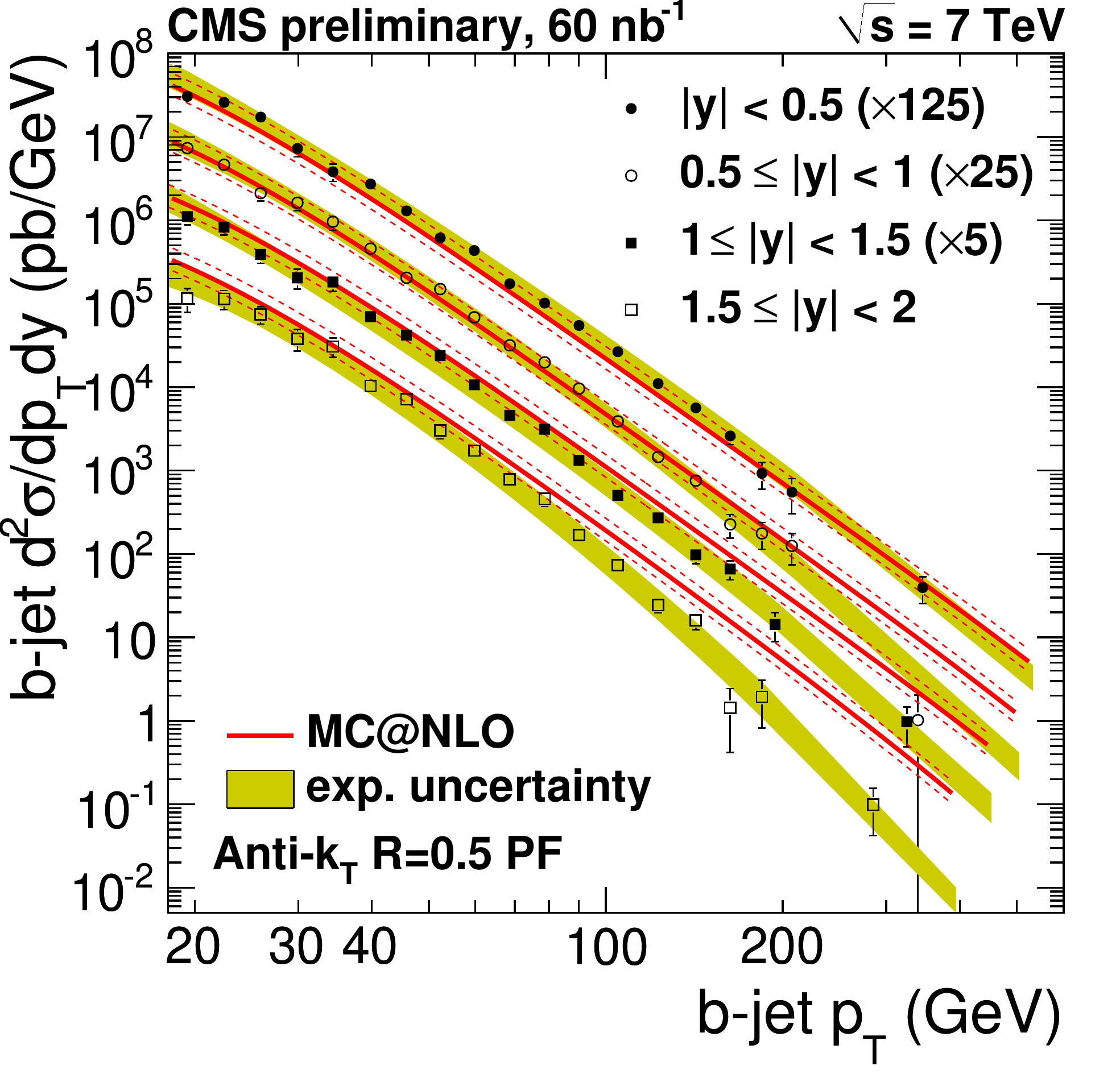}
\caption{Inclusive $b$-jet production cross section measured as a function of 
muon \pt\ from semi-muonic decays (left) and as a function of $b$-jet \pt\ 
from tagged jets (right) compared to simulation.
\label{fig:incl}}
\end{figure}

The inclusive $b$-quark production cross section is obtained by correcting the measured
$b$-quark yield by the selection efficiency in bins of muon \pt\ as shown in 
Figure~\ref{fig:incl}. The total
visible cross section is measured to be ($1.32\pm0.01 (\textrm{stat.}) \pm0.30 
(\textrm{syst.}) \pm0.15 (\textrm{lumi.})$) 
$\mu$b for $b$-jet
decays with a muon with $\pt > 6$ GeV and $|\eta| < 2.1$. The 
measured cross section is larger than that predicted by MC@NLO 
($0.95^{+0.42}_{-0.21}$ $\mu$b) and smaller than that predicted by Pythia (1.9 $\mu$b).

The second technique used to measure the inclusive beauty production rate
relies on the identification of displaced secondary vertices within reconstructed
jets to tag them as $b$ jets~\cite{btag}\hspace{-.35em}. An inclusive jet sample
is used to search for jets containing a secondary vertex. The secondary vertex is
required to contain at least three charged tracks and the vertex must be
well separated from the primary event vertex since long-lived $B$ hadrons give
rise to a larger separation than lighter jets. 
A separation cut is chosen such that $\approx 60\%$
efficiency is obtained with $\approx 0.1\%$ rate for mistagging light jets as $b$ jets.

The production of $b$ jets is calculated as a double differential cross section versus
jet \pt\ and $y$, where the reconstructed values have been corrected to the particle
\pt\ and $y$. The measured results are shown in Figure~\ref{fig:incl}.
The leading systematic uncertainties arise from the $b$-jet energy scale corrections,
the data-driven uncertainties on the $b$-tagging efficiency and from the mistag rates
for light jets. The overall agreement with MC@NLO is reasonable, though the
modeling of the rapidity dependence shows discrepancies between the data and simulation.

\section{$b\bar{b}$ correlations}

The secondary vertex finding technique also allows for the study of correlations
between $b\bar{b}$ pairs~\cite{cor}\hspace{-.35em}. 
The correlations between two $B$ candidates can provide
useful information about the $b\bar{b}$ pair production mechanism, where pairs produced from
gluon splitting are expected to have small separations, while those from flavor creation
are expected to dominate at large separation. Secondary vertices
are reconstructed with at least three charged tracks and a 3D flight length from the primary
vertex
greater than five times its uncertainty. The flight length of the $B$ candidate is computed as the direction
from the primary vertex to the secondary vertex. For events with exactly two such identified 
secondary vertices, the quantity $\Delta R = \sqrt{\Delta\phi^2 + \Delta\eta^2}$ is computed,
where $\Delta\phi$ is the difference in polar angle and $\Delta\eta$ is the difference in
pseudorapidity between the two $B$ candidate directions. 

\begin{figure}
\hspace{0.03\textwidth}
\includegraphics[width=0.42\textwidth]{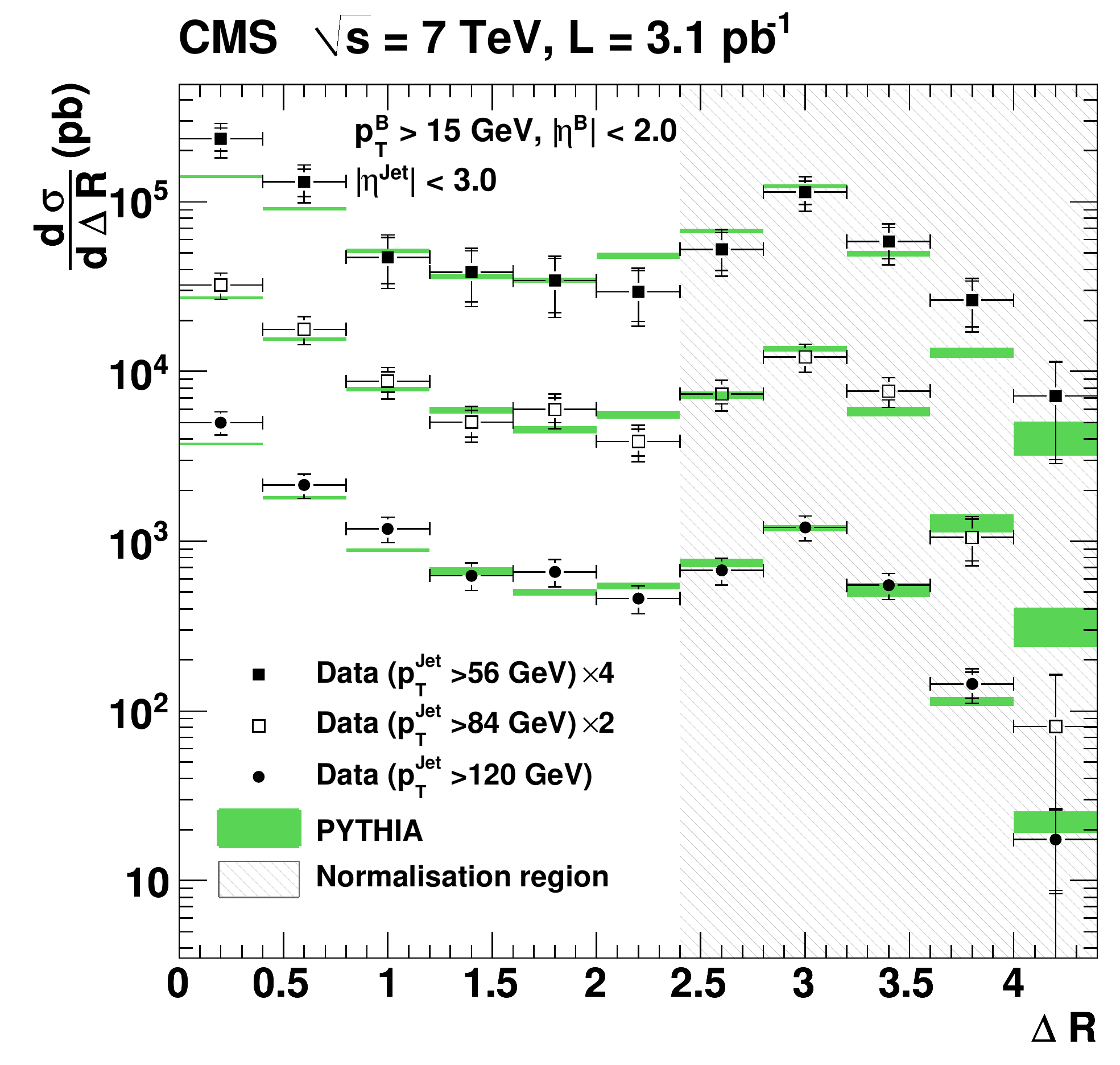}
\hspace{0.1\textwidth}
\includegraphics[width=0.42\textwidth]{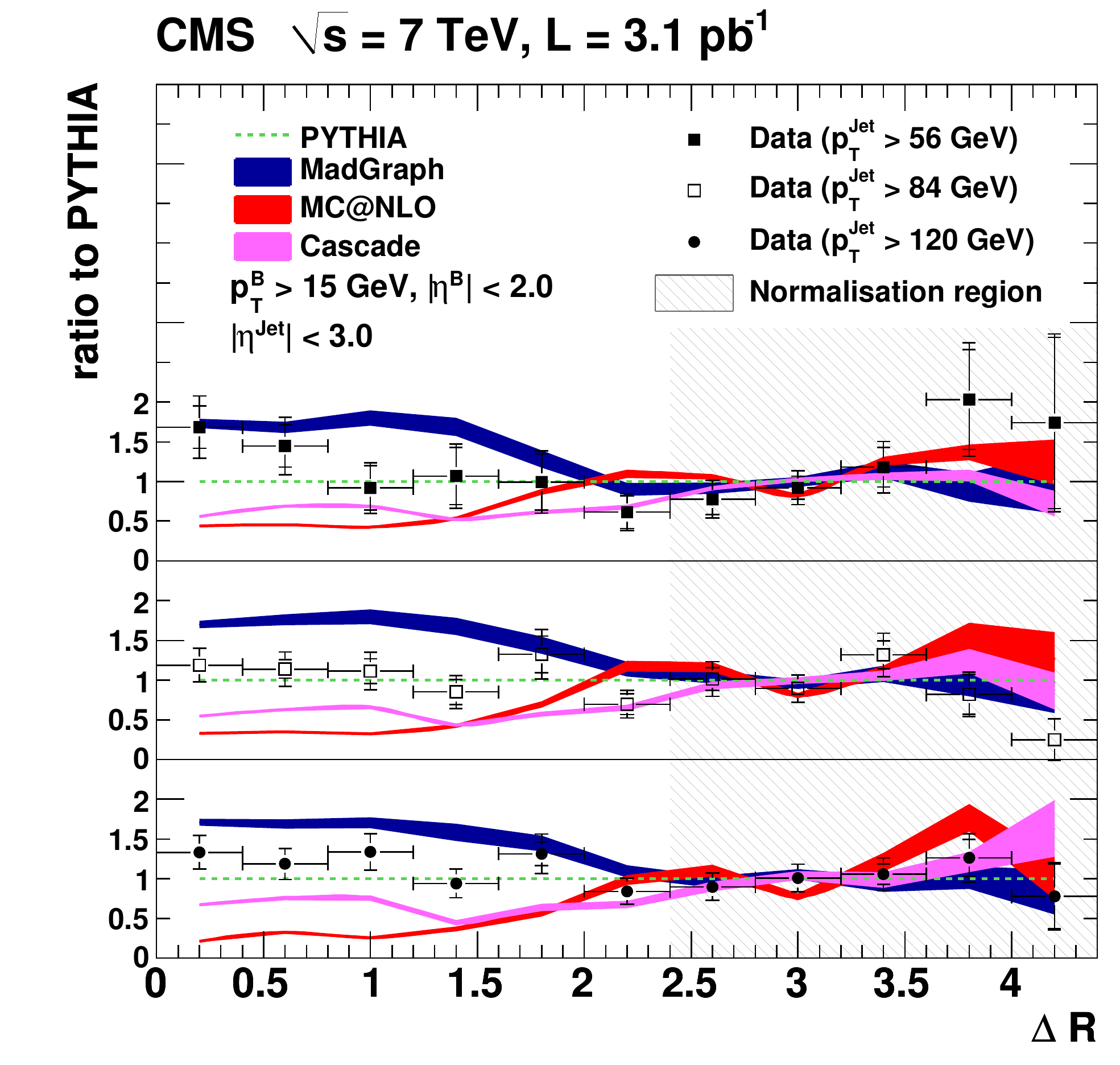}
\caption{Correlation results between reconstructed pairs of $b$-tagged jets compared
to Pythia simulation (left) and other theoretical predictions normalized to the 
Pythia result (right).
\label{fig:cor}}
\end{figure}

Results are shown in Figure~\ref{fig:cor} for events where both $B$ candidates have 
$\pt > 15$ GeV and $|\eta| < 2.0$. The reconstructed jet momentum is corrected back to
the true value, and the results are reported for three different regions of leading
jet \pt. The results are normalized to the region with $\Delta R > 2.4$, which is 
expected to be better understood theoretically. The data show 
an excess at low $\Delta R$
values compared to the prediction from Pythia suggesting that the contribution from
gluon splitting is larger than expected.

\section{Exclusive $B$ production}

A fully exclusive reconstruction technique is used to measure the cross sections for
$B^+$~\cite{Bp}, $B^0$~\cite{Bz}, and $B^0_s$ mesons. The three species are reconstructed
by fitting to a common vertex a $\JPsi$ plus a $K^+$, $K^0_S$, or a $\phi$, respectively. The 
$\JPsi$ mesons are reconstructed in their decays to $\mu^+\mu^-$, while the $K^0_S$ and
$\phi$ mesons are reconstructed in their decays to $\pi^+\pi^-$ and $K^+K^-$, respectively.
The dominant backgrounds in each analysis arise from events with a 
prompt $\JPsi$. To distinguish the signals
from these backgrounds, a two-dimensional fit to the $B$ mass and $B$ lifetime is used for
each $B$ species to extract the signal yield in bins of $B$ \pt\ and $y$. The fitted 
lifetimes in all three cases are consistent with the known values.

\begin{figure}
\includegraphics[width=0.47\textwidth]{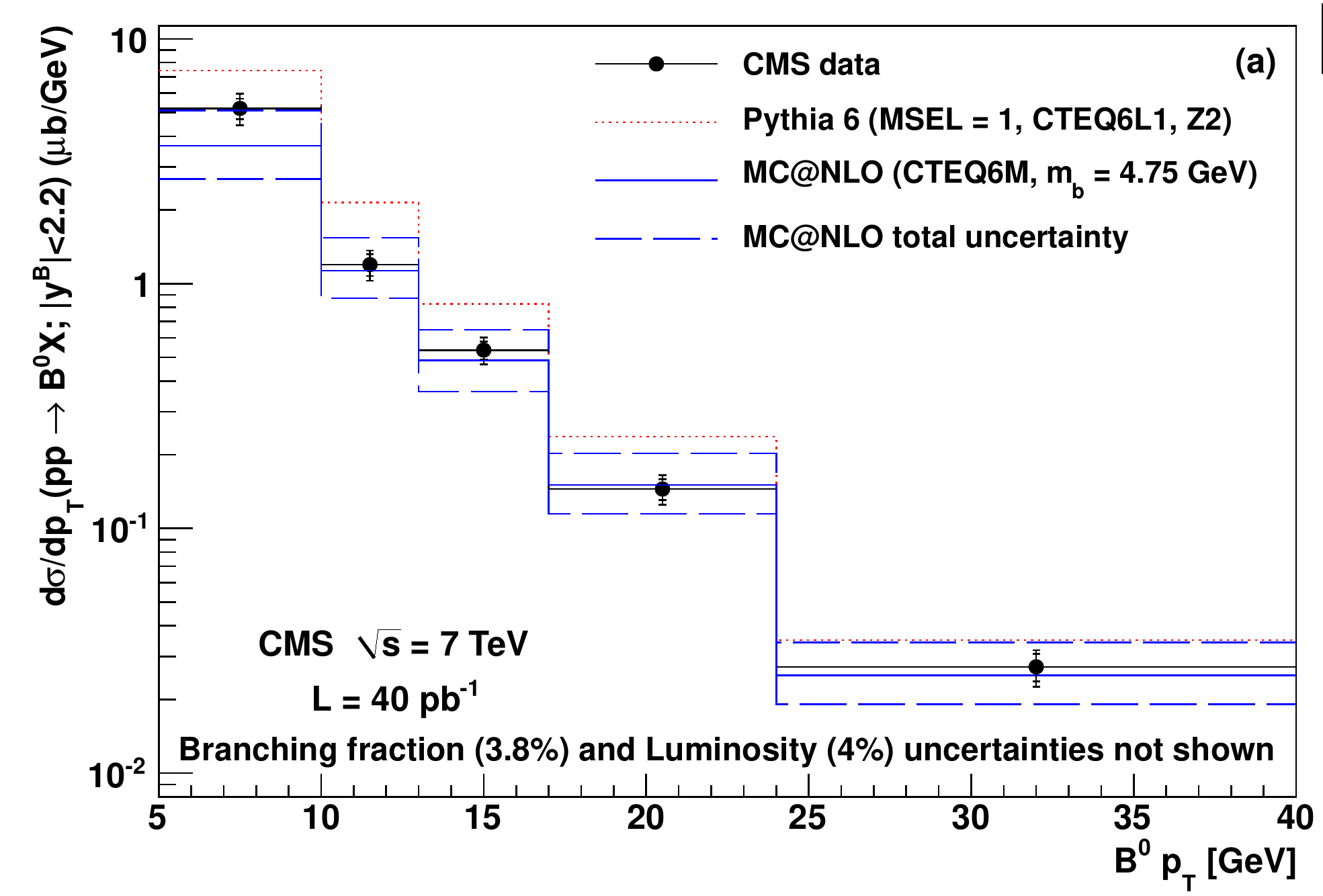}
\includegraphics[angle=90,width=0.52\textwidth]{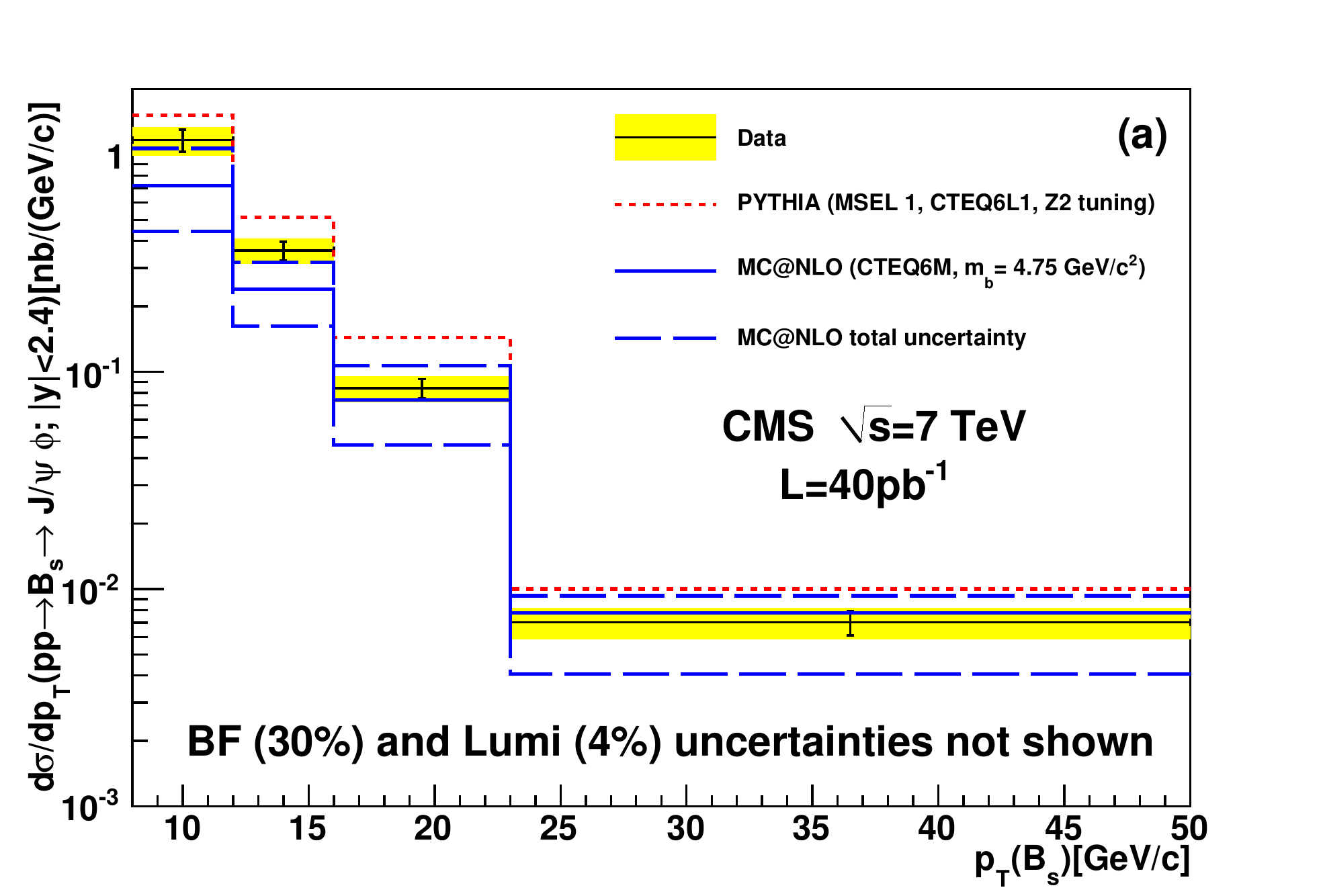}
\caption{Measured differential production cross sections versus \pt\ for $B^0$ (left) 
and $B^0_s$ (right) mesons compared to theoretical predictions.
\label{fig:exb}}
\end{figure}

For $B^+$, 912 signal events are observed in 6 $\textrm{pb}^{-1}$ of data, while 809 and 549 events 
are observed for $B^0$ and $B^0_s$ in 40 $\textrm{pb}^{-1}$.
The fitted signal yields are corrected for the detector acceptance and reconstruction and
trigger inefficiencies to calculate the cross section. For the $B^+$ and $B^0$ measurements,
candidates with $B$ $\pt > 5$ GeV are used, while for $B^0_s$ $\pt > 8$ GeV are considered.
For $B^+$ and $B^0_s$ ($B^0$) $B$ candidates are required to have $|y| < 2.4$ (2.2). The 
total visible cross sections are measured to be ($28.1\pm2.4\pm2.0\pm3.1$) $\mu$b for $B^+$ and
($33.2\pm2.5\pm3.5$) $\mu$b for $B^0$, where the first error is statistical, the second is systematic,
and the third for $B^+$ is the uncertainty in the luminosity, while the luminosity uncertainty is 
included in the systematic for $B^0$. The total visible cross section times the branching fraction
for $B^0_s\rightarrow\JPsi\phi$ is measured to be ($6.9\pm0.6\pm0.5\pm0.3$) nb for $B^0_s$. In
all cases, the observed cross sections are found to be lower than those predicted by Pythia
and higher than those predicted by MC@NLO, though compatible within uncertainties. The results
for $B^0$ and $B^0_s$ versus \pt\ are shown in Figure~\ref{fig:exb}.

\section{Conclusions}

A variety of measurements of heavy flavor production have been made by CMS
in $pp$ collisions at $\sqrt{s} = 7$ TeV. These include
$\JPsi$ and $\Upsilon$ double-differential production cross sections, 
measurements of inclusive beauty production from multiple complementary methods, 
including $b\bar{b}$ correlation measurements, and three
exclusive $B$ production cross section measurements. 
While the agreement with MC models is generally good,
none of the theoretical models considered yet describe
all of the features observed in the data.

\end{document}